\documentclass[aip,apl,reprint]{revtex4-1}
\usepackage{graphicx}
\pdfoutput=1
\begin{document}

\title{Photon noise limited radiation detection with lens-antenna coupled Microwave Kinetic Inductance Detectors}
\date{\today}

\author{S. J. C. Yates}
\email[e-mail: ]{s.yates@sron.nl}
\affiliation{SRON, Sorbonnelaan 2, 3584 CA Utrecht, The Netherlands}
\author{J. J. A. Baselmans}
\affiliation{SRON, Sorbonnelaan 2, 3584 CA Utrecht, The Netherlands}
\author{A. Endo}
\affiliation{Kavli Institute of NanoScience, Faculty of Applied Sciences, Delft University of Technology, Lorentzweg 1, 2628 CJ Delft, The Netherlands}
\author{R. M. J. Janssen}
\affiliation{Kavli Institute of NanoScience, Faculty of Applied Sciences, Delft University of Technology, Lorentzweg 1, 2628 CJ Delft, The Netherlands}
\author{L. Ferrari}
\affiliation{SRON, Landleven 12, 9747 AD Groningen, The Netherlands}
\author{P. Diener}
\affiliation{SRON, Sorbonnelaan 2, 3584 CA Utrecht, The Netherlands}
\author{A. M. Baryshev}
\affiliation{SRON, Landleven 12, 9747 AD Groningen, The Netherlands}
\affiliation{Kapteyn Astronomical Institute, University of Groningen, P.O. Box 800, 9700 AV Groningen, The Netherlands}

\begin{abstract}
Microwave Kinetic Inductance Detectors (MKIDs) have shown great potential for sub-mm instrumentation because of the high scalability of the technology. 
Here we demonstrate for the first time in the sub-mm band (0.1\ldots2~mm) a photon noise limited performance of a small antenna coupled MKID detector array and we describe the relation between photon noise and MKID intrinsic generation-recombination noise.
Additionally we use the observed photon noise to measure the optical efficiency of detectors to be $0.8\pm 0.2$.

\end{abstract}
\maketitle 

Advances in sub-mm astronomy (100\ldots1000~GHz) have always been driven by advances in radiation detection technology. Current state of the art imaging arrays consist of up to 1000~pixels of cryogenic detectors with a sensitivity approaching the photon noise limit~\cite{ACT}. 
Significant further advances in observing speed are only possible by increasing the pixel number. 
Microwave Kinetic Inductance detectors (MKIDs)~\cite{Day03} are very promising to use for very large arrays due to their intrinsic multiplexing capability~\cite{yatesAPL09}.
%Several telescope engineering runs have demonstrated increasing technological maturity~\cite{NIKA, CALTECH}. 
Despite the very low reported values~\cite{Pieter,Leduc10} of the (dark) Noise Equivalent Power (NEP) $\sim 3 \times 10^{-19}$W/Hz$^{1/2}$, MKIDs have yet to demonstrate photon noise limited performance.
In this letter we demonstrate photon noise limited performance of a lens-antenna coupled MKID array. The presented device geometry ensures that all radiation to the antenna is converted into quasiparticles in the sensitive part of the MKID, which is made of Al. This design, combined with the recent observation of generation-recombination noise in an Al MKID~\cite{Pieter}, enables the demonstration of photon noise limited radiation detection.

The fundamental limit of any ideal photon integrating sub-mm detector is the noise associated with the Bose-Einstein fluctuations in photon arrival rate~\cite{Boyd82}  which results in a photon noise limited NEP:
\begin{equation}
NEP_{photon}  = \sqrt{2PhF(1+mB)}\simeq \sqrt{2P hF}\label{eqn:photon} \\
\end{equation}
where $P$ is the sky power loading the detector at a frequency $F$ from the sub-mm source. $B$ is the photon occupation number per mode and $m$ the efficiency from emission to detection of one mode. The $(1+mB)$ term is the correction to Poisson statistics due to wave bunching~\cite{Boyd82}. In the presented experiment $m$ is small to enable low power coupling.
%, so while the full form of Eq.~\ref{eqn:photon} is used the approximation $1+mB=1$ gives an almost identical result.

MKIDs are superconducting pair breaking detectors, in which incident sky radiation changes the equilibrium between Cooper pairs and quasiparticles and hence the microwave conductivity in a planar superconducting resonator~\cite{Day03}. The number fluctuations between quasiparticles and Cooper pairs in the MKID will limit the device sensitivity giving a generation-recombination (g-r) NEP~\cite{Wilson01, Pieter}:
\begin{equation}
NEP_{g-r}  = \frac{2\Delta}{\eta_{pb}}\sqrt{N_{qp}/\tau} 
\end{equation}
In the limit where the quasiparticle number in the MKID is dominated by the sky power then NEP$_{G-R}$ can be approximated by:
\begin{equation}
NEP_{g-r} = \sqrt{2 P \Delta/\eta_{pb}}  \label{eqn:g-r}
\end{equation}
where we only consider recombination noise contributions as quasiparticle creation is correlated with the photon arrival.
Here $N_{qp}$ is the number of quasiparticles, $\tau$ is the quasiparticle recombination lifetime, $\eta_{pb}\sim 0.6$ is the efficiency for conversion of energy into quasiparticles~\cite{Kozorezov00} and $\Delta$ is the superconducting gap.
Under illumination we therefore find that the ratio $NEP_{photon}/NEP_{g-r}\simeq \sqrt{hF\eta_{pb}/(\Delta)}$.
Note that for MKIDs the detector speed is set by the quasiparticle lifetime. 
Hence any signal in quasiparticle number, such as a sky signal or quasiparticle noise signal, has a Lorentzian roll-off in the measured noise spectra and signal response of $1/\sqrt{1+(2\pi f \tau)^{2}}$ with $f$ the signal modulation frequency~\cite{BarendsPRL08,Pieter}. 
Therefore photon noise limited detection in a MKID is associated with a white noise spectrum, rolled-off at the quasiparticle lifetime and with a $NEP\propto \sqrt{P}$ at a level a higher than $NEP_{photon}$ due to the g-r noise contribution.
MKIDs also have non-fundamental noise contributions. 
Using phase readout~\cite{Day03}, MKIDs are limited by two level system fluctuations in the dielectric which have a $f^{-0.25}$ or steeper spectral contribution to the $NEP$ rolled-off by the resonator ring time~\cite{Gao07}. In the MKID amplitude (or dissipation) readout~\cite{Gao07,Jochem08}, MKIDs are typically limited by the noise contribution of the amplifier chain, even down to the quantum noise level~\cite{Gaopre10}.

\begin{figure}
\includegraphics*{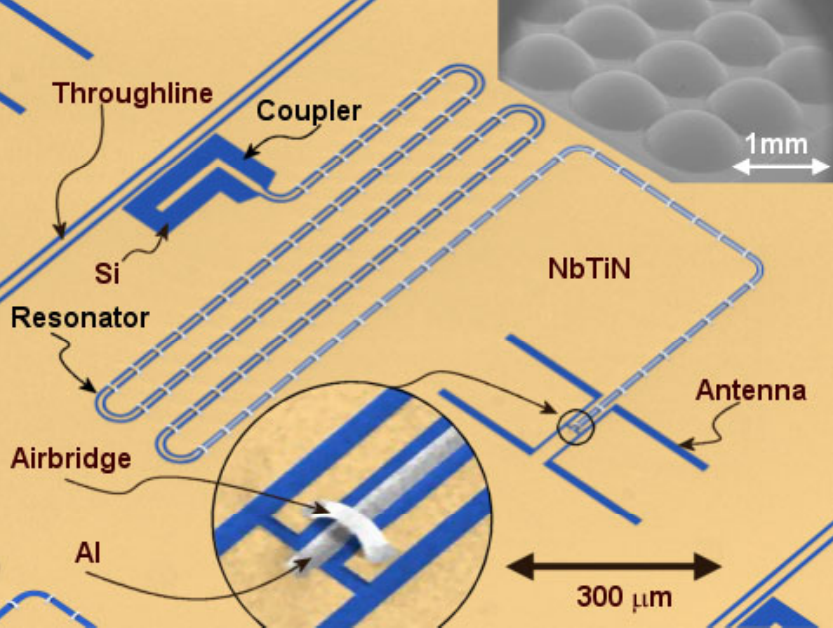}
\caption{(Color online) scanning electron micrograph (SEM) of the hybrid MKIDs used. Indicated is NbTiN (gold), the silicon substrate (blue) and aluminum (gray) which is used as the sensitive area of detector and for airbridges. Inset bottom shows the aluminum airbridges. Inset top right is a SEM of a laser machined Si lens array, used to couple radiation via the antenna into the MKID.}\label{fig:device}
\end{figure} 

In the experiment presented we study an array of 8$\times$9 pixels of lens-antenna coupled MKIDs.
The MKIDs are quarter wavelength coplanar waveguide (CPW) superconducting resonators~\cite{Day03} weakly coupled to a CPW line (throughline, with a 6~$\mu$m wide gap with 10~$\mu$m central line) which connects all pixels (see Fig.~\ref{fig:device}). 
The resonator acts as a short for a probe signal on the throughline at a frequency $F_0\sim$~6~GHz. 
At its shorted end the MKID CPW acts as the feed to a single polarization sensitive twinslot antenna~\cite{Filipovic93}, optimized for 325 GHz radiation, placed in the geometrical focus of a Si lenslet mounted to the back of the detector chip. 
The lenslet is part of a laser machined Si lens array as shown in the insert of Fig.~\ref{fig:device}. 
The only affect the antenna has on the MKID properties is a slight shift in MKID resonance frequency.
The devices~\cite{Akira} are made of 300~nm sputtered NbTiN, a superconductor~\cite{NbTiNprop} with $2\Delta= 1.1$~THz and so lossless at both probe and sky frequency.
The central line of the MKID close to the antenna is made of sputtered 80~nm thick Al, $2\Delta=90$~GHz and normal state resistivity 1.1~$\mu\Omega$cm, for a length of 1.2 mm. 
Hence radiation with $1.1$~THz$>F>90$~GHz traveling from the antenna into the resonator will be absorbed \emph{only} in the Al central line.
The created quasiparticles are confined within the Al because of the presence of an Andreev barrier due to the difference in superconducting gap between Al and NbTiN.
To prevent parasitic odd modes at the sky frequency coupling to the MKID~\cite{Neto}, lithographical airbridges of 200~nm sputtered Al are manufactured using a resist reflow technique over the MKID CPW.
The MKID CPW has a 4.5~$\mu$m wide gap and central line, both narrowing to 3~$\mu$m near the antenna while the coupler has a 30~$\mu$m gap with a 10~$\mu$m central line.

The arrays are tested at 100~mK in a commercial adiabatic demagnetization refrigerator with a cryogenic blackbody radiator as calibration load.
The sample holder containing the detector chip and Si lens array is placed into a closed box at 100~mK. 
Electrical connection for readout from the outside of the cryostat to the sample holder is done via two coaxial cables and one Low Noise Amplifier (LNA) at 4~K with a $\sim$~4K noise temperature.
The variable temperature blackbody radiator, a 40~mm diameter copper cone, is placed in a separate 4~K box opened with a 1~cm aperture and a 1~THz lowpass filter. 
Radiation coupling from sample to blackbody is defined by placing on the 100~mK box an aperture of 2mm with a bandpass filter of 90~GHz around 325~GHz. 
To obtain high emissivity for radiation emission and stray light absorption a coating~\cite{SPIREblack,HIFIblack} of carbon loaded epoxy and 1 mm SiC grains is used on the inside of the 100~mK box and blackbody cone.
Power radiated on each pixel is numerically calculated from the blackbody temperature by integrating the Planck equations with one polarization versus frequency for the optical throughput over the measured filter transmission.

\begin{figure}
\includegraphics*{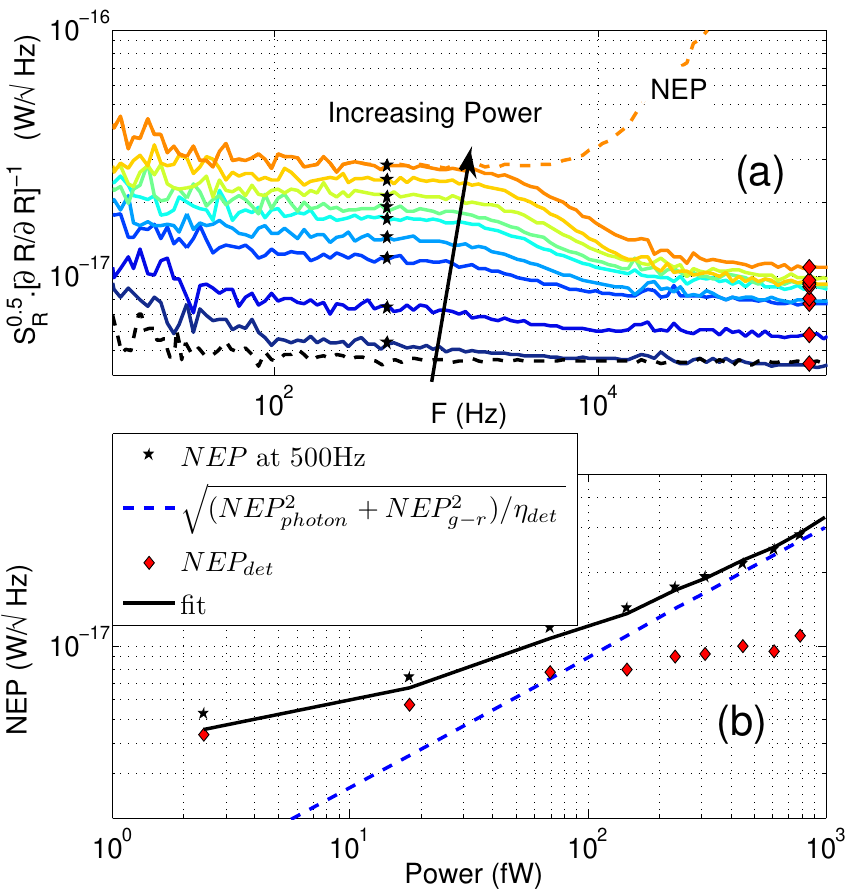}
\caption{(Color online) (a) Optical spectral density (in W/Hz$^{1/2}$) in MKID radius readout under black body illumination. Higher curves are at higher power. The quasiparticle roll-off is visible above 1~kHz, below which this is identical to the $NEP$. The higher dotted curve is the highest power optical $NEP$ including the quasiparticle roll-off. The lower dotted curve is the LNA noise floor for the lowest power.
(b) 500Hz $NEP$ value versus black body power. The stars are the measured 500Hz $NEP$ and the diamonds $NEP_{det}$ which are also indicated in (a). The dotted line is the photon and g-r noise contribution, while the solid line is the fit to the measured data for the detector optical efficiency, plotted for $\eta$=0.8.}\label{fig:photonnoise}
\end{figure} 

The detector optical NEP is measured at various loading powers by measuring the detector noise and responsivity at various temperatures of the blackbody calibration load. 
We use conventional MKID phase and MKID amplitude readout (see~\cite{Jochem08}) based on the homodyne detection of the complex transmission of a probe signal at the MKID resonance frequency $F_{0}=6.301$~GHz. 
Here we present MKID amplitude readout because then the only additional noise source which is present on top of the photon and g-r noise is the one from the LNA, which can be directly measured. 
The amplitude signal, denoted by R, is measured as a function of time for a stable blackbody reference temperature ($T_{ref}$), from which the MKID amplitude power spectral density ($S_R$) is calculated. 
Subsequently the amplitude signal ($R(T)$) is measured while slowly varying the blackbody temperature over a small range around $T_{ref}$, from which we obtain the optical responsivity $\partial R/\partial P$ by a linear fit to the data.
The experimentally obtained optical $NEP$ is then:
\begin{equation}
NEP=\sqrt{S_R} \cdot \left(\frac{\partial R}{\partial P}\right)^{-1} \sqrt{1+(2\pi f \tau)^2}
\end{equation}
This procedure is repeated at various blackbody reference temperatures representing different simulated sky background loading powers\cite{note:Qfactor,note:pdis}.
In Fig.~\ref{fig:photonnoise}a) we show $S_R.[\partial R/\partial P]^{-1}$ for all powers which gives the spectral shape of the noise spectrum and is equal to the $NEP$ at modulation frequencies $f<1/(2\pi\tau)$. 
For the lowest powers the spectra are dominated by the white noise contribution of the LNA. 
At higher powers a white noise spectrum with a loading power dependent roll-off frequency $\gtrsim$~1.5~kHz increasingly dominates, consistent with a separate direct measurement of the lifetime ($\sim$150~$\mu$s) using a decay response to a LED pulse~\cite{Jochem08,BarendsPRL08}.
Fig.~\ref{fig:photonnoise}b) (black stars) shows the measured optical $NEP$ versus power, taken at 500~Hz to avoid $1/f$ noise due to the readout chain and drift in the blackbody temperature. 
The measured $NEP$ has a $\sqrt{P}$ dependence at high power which in combination with the spectral shape of the noise is the signature of photon noise limited performance.
Under a loading of $P\sim 2$~fW the optical $NEP\sim~5~\times 10^{-18}$W/Hz$^{1/2}$, dominated by the LNA. 
The photon and g-r noise dominate at $P>$~100~fW up to $P$=20~pW~\cite{notepower}. 
Sky loading power for ground based imaging arrays is typically of the order of 1\ldots20~pW, hence the measured array is photon noise limited for these applications except at low modulation frequencies due to 1/f noise contributions. 

The measured optical $NEP$ can be described by:
\begin{equation}
NEP^{2}=NEP^{2}_{det}+(NEP^{2}_{g-r}+NEP^{2}_{photon})/\eta \label{eqn:sum}
\end{equation}
Here $\eta$ is the optical efficiency of the detector, defined as the ratio of the power in front of the lens and the power detected~\cite{Benford98,note:detNEP}.
$NEP_{det}$ is the extra contribution from the detector or readout. 
For MKID amplitude readout $NEP_{det}$ is determined by the noise contribution from the LNA, shown in Fig.~\ref{fig:photonnoise}a) by the lower dotted line as being independent of frequency.
At modulation frequencies $f\gg 1/(2\pi\tau)$ then $S_R$ is equal to the LNA noise floor.
Hence $NEP_{det}$ at all modulation frequencies is given by the value of $S_R.[\partial R/\partial P]^{-1}$ at 200~kHz.
Therefore Eq.~\ref{eqn:sum} enables us to find $\eta$ using the measured $NEP_{det}$ and the numerically calculated values of $NEP_{photon}$ and $NEP_{g-r}$. We find $\eta$=~0.8$\pm$0.2 for a single polarization~\cite{note:error}.
Other experiments using different arrays, bath temperatures and loading powers give reproducible results.
The measured optical efficiency is in agreement with both a direct 3D electromagnetic simulation~\cite{Lorenza} of the lens-antenna used in the experiment and older, more general calculations~\cite{Filipovic93}.

In conclusion, we have demonstrated high optical efficiency photon noise limited detection with a lens-antenna coupled MKID array made of Al and NbTiN at loading powers above 100~fW. This opens the way to the planning and building of the next generation of very large cameras for sub-mm astronomy based on MKIDs. We also demonstrate that the devices can be calibrated using a measurement of the photon noise. This is relevant for MKIDs as they are not direct power meters like bolometers. 

The authors would like to thank Andrea Neto, Giampiero Gerini, Marcel Bruijn, Jan-Joost Lankwarden, Alisa Haba, Simon Doyle, Peter Ade and Carole Tucker for help with this work. Akira Endo is supported by NWO (Veni grant 639.041.023) and JSPS (Fellowship for Research Abroad 215). This work was in part supported by ERC starting grant ERC-2009-StG Grant 240602 TFPA.

%\newpage

\begin{thebibliography}{00}
\bibitem{NbTiNprop} NbTiN has a Tc of 14K and normal state resistivity of 170~$\mu\Omega$cm. From the downshift between design and measured resonant frequency (f/fdes=0.8) for an all NbTiN MKID, we estimate the kinetic inductance fraction $\alpha$ for NbTiN at 0.35.
\bibitem{note:Qfactor} At low radiation loading under a readout power of -76~dBm (-31~dBm internal power~\cite{Gao07}) the MKID has a Q factor~\cite{Day03} of 58000, a coupling Q-factor (Q$_{\mathrm{c}}$) of 61000, an internal Q-factor (Q$_{\mathrm{i}}$) of $1\times10^6$. At 1~pW optical loading Qi goes to $2\times10^5$ for a readout power of -72dBm (-29~dBm internal power). The kinetic inductance fraction of the Al section calculated using the method in reference~\cite{BarendsThesis} is 5~\%.
\bibitem{note:pdis} The microwave readout will dissipate $\sim$~2 to 20~pW from lowest to highest optical loading. Note that the readout current is dissipated in the quasiparticles, so only capable of pair breaking indirectly.
\bibitem{notepower} The range from 1\ldots20~pW was measured measured in a separate experiment with a large blackbody aperture.
\bibitem{note:detNEP} $NEP_{det}$ is a taken from the measured data, so is not renormalized by $\eta$.
\bibitem{note:error} A careful analysis of potential sources of systematic error, backed by laboratory measurements of individual optical components, indicates that the statistical errors dominate and are hence the limits presented.
\end{thebibliography}
\end{document}